# Effect of electrolyte concentration and symmetry on the heterogeneous surface charge in an electrically gated nanochannel

Movaffaq Kateb[a,b,*,†], Morteza Fathipour[b], Mohammadreza Kolahdouz[b,*]

[a] BIOS lab on chip, MESA+ Institute for Nanotechnology, University of Twente, Enschede, Netherlands
[b] School of Electrical and Computer Engineering, University of Tehran, Tehran, Iran

**Abstract:** The present study aims to investigate utilizing field-effect for inducing heterogeneous surface charge and consequently changing the fluid flow in a solid-state nanochannel with converging-diverging periodicity. It is shown that the combination of geometry and applied gate voltage ($V_G$) would generate heterogeneous surface charge at the channel walls which can be modulated by $V_G$, i.e. a moderate $V_G$ (0.7-0.9 V) causes charge inversion in diverging sections of the channel ($D_{max}$) while $V_G > 0.9$ enables charge inversion in the entire channel but it is still non-uniform in each section. The results show that zeta ($\zeta$) potential is a function of $V_G$ which shows a linear to non-linear transition due to dilution of electrolyte in agreement with density functional theory and Monte Carlo simulations. In contrast, electrolyte symmetry has a minor effect on the variation of $\zeta$ potential. It is also shown that the difference in $\zeta$ potential across the channel ($\Delta\zeta$) increases by dilution of electrolyte and utilizing a more symmetric electrolyte with lower valances. For the first time, it is shown that $\Delta\zeta$ presents a maximum with the $V_G$. The $V_G$ corresponding to the maximum $\Delta\zeta$ decreases with both dilution of electrolyte and higher anion valance. This is of practical importance to overcome leakage current problem of field-effect fluidic devices. It is also shown that the velocity field can be altered by changing both electrolyte concentration and symmetry. However, applying $V_G$ was found to be a more efficient way than electrolyte modifications. This includes generating circulation inside the channel which is of prime importance for applications such as mixing or separation/trapping.

## 1. Introduction

Among the variety of force fields applied to manipulate ions, fluids, and particles, electrokinetic transport is the most popular and versatile one. This is mainly because electric fields can be easily scaled down to micro/nanoscale [1, 2]. However, the electrokinetic transport phenomena strongly depend on the inherent or induced zeta potential of the capillary/particle. The intrinsic zeta ($\zeta$) potential depends on, among other things, the nature of the solid surface, the concentration of ions, and the pH of the electrolyte. Thus, enormous efforts have been put to modify the channel walls and consequently electrokinetic transport permanently or temporarily. On the other hand, reduction of the flowrate was attributed to defects or adsorption of analyte on the channel walls and thus any modification was considered to be undesirable [3]. This controversy motivated numerous theoretical studies to investigate non-uniform $\zeta$ potential.

Anderson and Idol [4] analytically studied electroosmotic flow in a microchannel of periodically varying $\zeta$ potential and found out a reverse electroosmotic flow in specific regions generates circulation which they held responsible for the observed reduction of the flow rate. Ajdari [5, 6] showed that a symmetric alternating $\zeta$ potential with zero average (i.e. ± $\zeta$) only can lead to circulation without a net flow rate. Thus, the reduced flow rates resulting from a non-uniform $\zeta$ potential can be attributed to the stirring which can be utilized in designing passive mixers [7]. To realize such a mixer, however, the tradeoff between mixing and transport has to be resolved i.e. excellent mixing may lead to poor transport efficiency and *vice versa* [8]. Furthermore, these studies [3-8] were suffering from assumptions such as the thin electrical double layer (EDL) and Boltzmann distribution of ions which only applies to a diluted aqueous solution in a microchannel. Yet another crucial assumption was neglecting ionic advection which does not satisfy in a mixer that employs circulation. Fu *et al.* [9] studied a step-change in $\zeta$ potential by determining ionic distribution perpendicular to the channel wall using Nernst-Planck equations as opposed to the classical Boltzmann model. It has been shown that the former correctly yields to a gradual change of EDL while the latter results in a sharp change around the $\zeta$ potential step, a conclusion which cannot be supported by the physics of the problem due to the convective transport of ions.

Despite all these great efforts, no resolution has yet been introduced for the low flow rate problem of non-uniform $\zeta$ potential [3]. To gain active control over $\zeta$ potential, Qian and Bau [10] proposed a field-effect transistor (FET) structure consisting of alternatively charged electrodes separated from the fluid by a very thin insulator. The principals of utilizing a perpendicular electric field to modulate $\zeta$ potential had already been discussed both analytically [11] and experimentally [12]. In the FET structure, a potential is applied to the gate electrode patterned on the outer surface of a dielectric channel wall. This allows generating an electric field in the dielectric channel walls as well as to change electrochemical potential within the EDL (over Debye length) through the capacitance. Thus, field-effect allows much more rapid control over electrokinetic transport e.g. it can be pulsed much faster than changing the pH. This consideration is practically important since the chemistry of analyte may restrict choices of the pH, the ionic concentration as well as the material used for the channel walls [11].

---

* Corresponding Authors: Email addresses: movaffaqk@ru.is, kolahdouz@ut.ac.ir
† Present address: School of Science and Engineering, Reykjavik University, Menntavegur 1, IS-102 Reykjavik, Iceland.

Recently, an enormous theoretical effort has been devoted to understanding nanofluidic devices implementing the FET concept, including regulation ions, fluid, and biomolecules transport [13-26]. However, these studies have several limiting assumptions, some by assuming constant surface charge density at the nanochannel wall [14-22], some considered just one (background) ionic species [13-26] and others neglecting the Stern layer effect [14-25]. Besides, these studies described field-effect within a uniform channel. Thus, developing a more general and realistic model to elaborate experimental observations in relevant gated nanofluidic devices with non-uniformity is highly desirable. Qian and Bau [10] studied a channel of non-uniform $\zeta$ potential and showed this FET overcomes the above-mentioned mixing and transport tradeoff by sequential mix and transport. However, the realization of such a device has not been reported so far perhaps because of challenges embedding multiple gates into a microfluidic channel. We have recently reported successful design and fabrication of a simple nanofluidic FET with non-uniform $\zeta$ potential which utilized a single gate [27]. It has been shown that utilizing field-effect in a nanochannel with a variable cross-section enables generating non-uniform $\zeta$ potential. It has been also shown that inverse charge can be obtained in such a device using relatively low gate potential (0.6-1.2 V). This achievement is of practical importance since most of the fabricated fluidic FETs reported to date require a very high gate voltage (9-50 V) [28, 29] which leads to significant leakage current [28].

In the present study, vertical (out-of-plane) channels in a large area were prepared by the BOSCH process, a deep reactive ion etching for fabrication of high aspect ratio structures. The sequential nature of the BOSCH process leaves scalloped sidewalls which are advantageous for generating heterogeneous $\zeta$ potential as will be discussed later. We have studied the behavior of this device under different conditions by simulation since out-of-plane channels do not allow imaging along the channel axis. This includes different concentrations of electrolyte as well as various electrolyte symmetries.

## 2. Method
### 2.1. Fabrication
As described in [27], the fabrication process began with the deposition of a 70 nm thick low-stress silicon nitride (SiRN) on a (100) p-silicon wafer. Then, several 2×2 mm² squares were patterned on the backside of the wafer using standard photolithography technique followed by immersion in a hot phosphoric acid bath to open a window in SiRN. Then, silicon was removed by KOH anisotropic wet etching through the window to fabricate suspended SiRN/Si diaphragm which became slightly transparent when Si thickness approached nearly 1 μm. The front face SiRN was then patterned by laser interference lithography [30]. In the next step, a hole array was made through SiRN using anisotropic etching in $SF_6$ and $C_4F_8$ mixture in an inductively coupled plasma system. The remaining SiRN serves as a hard mask during Si etching. The BOSCH process consists of 1, 1, and 2 s sequence of $C_4F_8$, $O_2$, and $SF_6$ at –40 °C, respectively. 1 kW power was utilized for drilling silicon through the SiRN mask. The backside again was dry-etched in a mixed ambient of $SF_6$, $O_2$, and $C_4F_8$ using 1 kW to open pore bottoms and reaching desired membrane thickness. Then, an oxide layer was grown everywhere by dry oxidation at 945 °C to reach the desired thickness.

### 2.2. The model
**Figure 1a** shows the scanning electron microscope image of the fabricated channel consisting of a periodic converging-diverging section. Figure 1b illustrates the geometry used in the numerical simulation for the channel walls and the gate were assumed to be made of $SiO_2$ and Si, respectively. The $Si/SiO_2$ interfacial region obtained in practice probably consisted of a thin $SiO_x$ layer which could act as a source for interfacial fixed oxide charges. In our numerical calculations, we have assumed that these charges can be neglected. The average diameter ($D_{ave}$) of the channel was assumed to be 130 nm which was smoothly changed between 100 and 160 nm in converging ($D_{min}$) and diverging ($D_{max}$) parts respectively, with a repetition period of 100 nm. Comparing figure1a and 1b, it can be seen the $D_{min}$ was sharper in the fabricated channel compared to the model used in the simulation. However, the fabricated channels with 200, 300, and 400 nm (not shown here) present smoother edge at $D_{min}$ and thus sharp edges were avoided in both gate and oxide for the simulation. The gate and oxide thicknesses are minimum at $D_{max}$ and maximum at $D_{min}$ in the model.

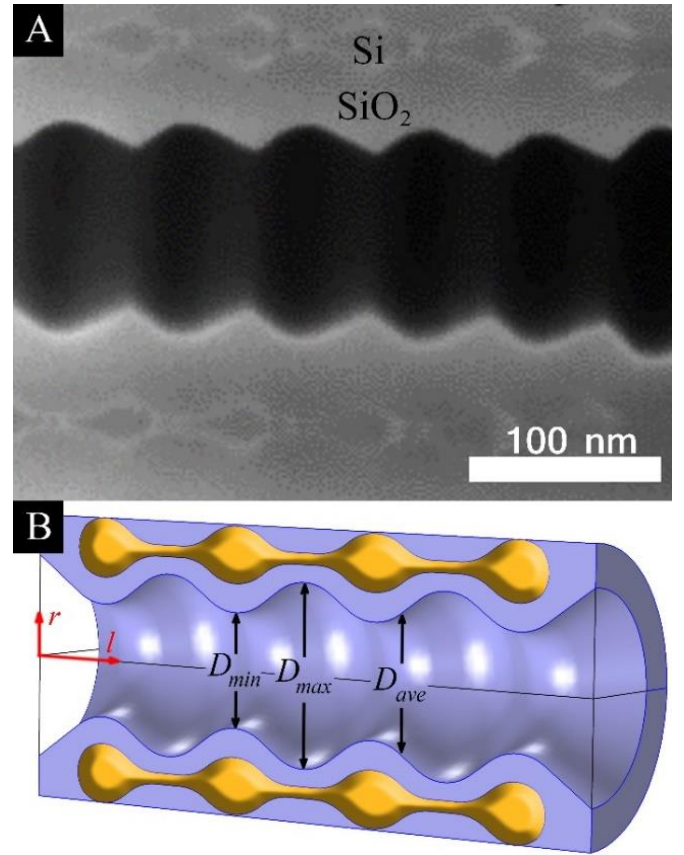

**Figure 1.** (a) SEM image of the channel cross-section obtained in the fabrication and (b) the geometry of channel in the simulation, which is also indicating longitudinal, $l$, and radial, $r$ axis.

### 2.3. Numerical approach
The EDL was modeled by solving mass transport of ions through diffusion, ion-ion interaction, and fluid convection as well as electro-migration due to the applied fields. These assumptions are necessary since the contribution of converging-diverging geometry and field-effect is still unknown and may alter charge distribution and EDL. The time-dependent form of the Nernst-Planck equation was used to describe the motion of ionic species inside the channel.

$$\frac{\partial c_i}{\partial t} + U \cdot \nabla c_i = \nabla \cdot \left[ D_i \nabla c_i + \frac{D_i z_i e}{k_B T} c_i \nabla \phi \right] \quad (1)$$

where $c_i$, $D_i$, and $z_i$ are the $i$ specie concentration, diffusion coefficient, and ionic valence, respectively. $U$ is the velocity vector of the fluid and $\phi$ is the electric potential. The $e$, $k_B$, and $T$ stand for the elementary charge, Boltzmann constant, and absolute temperature, respectively.

The mean-field approximation of the electrostatic potential is described by the Poisson relation, that relates the electrical potential to the charge density.

$$\nabla^2 \phi = -\frac{\rho_{ch}}{\varepsilon} = -\frac{e}{\varepsilon} \sum_{i=1}^{n} z_i c_i \quad (2)$$

here $\rho_{ch}$ and $\varepsilon$ are the charge density and electric permittivity with $n$ being the total number of species in the system. We assumed similar $\varepsilon$ for the bulk electrolyte and EDL and neglected its variation with $V_G$ which the latter is a valid assumption in non-overlapping EDL [31].

The Navier-Stokes equation is an expression of conservation of linear momentum for an incompressible Newtonian fluid with constant mass density. Now, we allow for fluid motion due to electrostatic force by adding a term to the Navier-Stokes to represent the body force density due to electrostatic force.

$$\rho_f \left( \frac{\partial U}{\partial t} + U \cdot \nabla U \right) = -\nabla p + \eta \nabla^2 U + \rho_{ch} E \quad (3)$$

here $U$, $\rho_f$, $\eta$, and $p$ are the velocity vector, density, viscosity, and pressure of the fluid, respectively. $E$ is the electric field due to charge redistribution around the walls and electric field applied along the channel as well as the radial field imposed by $V_G$. An electric field of $10^5$ V/m was generated along the channel by applying proper voltage to the inlet with respect to the outlet to generate electroosmotic flow.

Although no pressure gradient was imposed along the channel, neglecting the pressure, $p$, term implies skipping osmotic pressure effect which is proportional to the charge density. The later changes abruptly next to channel walls in the present geometry while neglecting $p$ assumes constant $\rho_{ch}$ everywhere in the channel. This is an important issue since it may lead to an artificial flow due to an unbalanced electro-hydrostatic ion pressure stemming from the Maxwell stress term.

The left-hand side of the equation (3) represents the convective transfer of linear momentum which can be simplified by neglecting unsteady terms. This is an appropriate assumption unless there is forcing at a high frequency such as shock wave. Besides, the continuity equation for an incompressible fluid requires that:

$$\nabla U = 0 \quad (4)$$

The later assumptions have been utilized for numerical simulation of heterogeneous $\zeta$ potential with [10] and without [32] field-effect (gate). For the case of non-uniform $\zeta$ potential, it has been demonstrated that the results of keeping [33] or skipping [9] the inertial term are in good agreement especially for ion distribution which determines EDL and electroosmotic flow.

The fluid was water with relative permittivity of 87.5 at room temperature, which is a proper solvent for most of the electrolytes. On each side of the channel, there is a reservoir with a constant concentration of electrolyte. We compared 100, 10, 1 and 0.1 mM concentrations of a monovalent electrolyte as well as 10 mM electrolytes of different symmetries i.e. $1^+$:$3^-$, $1^+$:$2^-$, $2^+$:$2^-$, $2^+$:$1^-$ and $3^+$:$1^-$.

The coupled system described above was simultaneously solved with a commercial finite-element package COMSOL (version 5.2) and Matlab R2016a. We also discussed a couple of issues in the existing experimental results that can be avoided using our device and compared the trend of our results to that of analytical and numerical simulations.

## 3. Results and discussion
### 3.1. Space charge density

**Figure 2** shows the variation of $\rho_{ch}$ at the channel walls with $V_G$ for different concentrations and electrolyte symmetries. The average (indicated by symbols) in each data set shows the $\rho_{ch}$ at $D_{ave}$, while, minimum and maximum are the $\rho_{ch}$ values at $D_{max}$ and $D_{min}$, respectively. At zero $V_G$, an increase in the concentration and cation valance results in higher positive $\rho_{ch}$ at the walls. Besides, none of the concentration and electrolyte symmetry used here were able to generate heterogeneous surface charge i.e. $\rho_{ch}$ is identical at $D_{max}$, $D_{min}$, and all channel walls. We call this value the intrinsic surface charge ($\sigma_0$). Note that an increase in $V_G$ increases heterogenous $\rho_{ch}$ at the walls as indicated by the width of shaded areas. Recently, we have shown that the origin of such a heterogeneous surface charge is a significant difference in the radial component of the electric field at $D_{max}$ and $D_{min}$ [27]. Thus, it is a combination of the channel geometry and the applied field that allows modulation of heterogeneous surface charge. By applying a proper $V_G$, it is possible to have zero charges at $D_{ave}$ accompanied by negative $\rho_{ch}$ at $D_{max}$ while $\rho_{ch}$ is still positive at $D_{min}$. This occurs at ~0.7 V for all cases and it shows a negligible variation by the change in the concentration and electrolyte symmetry.

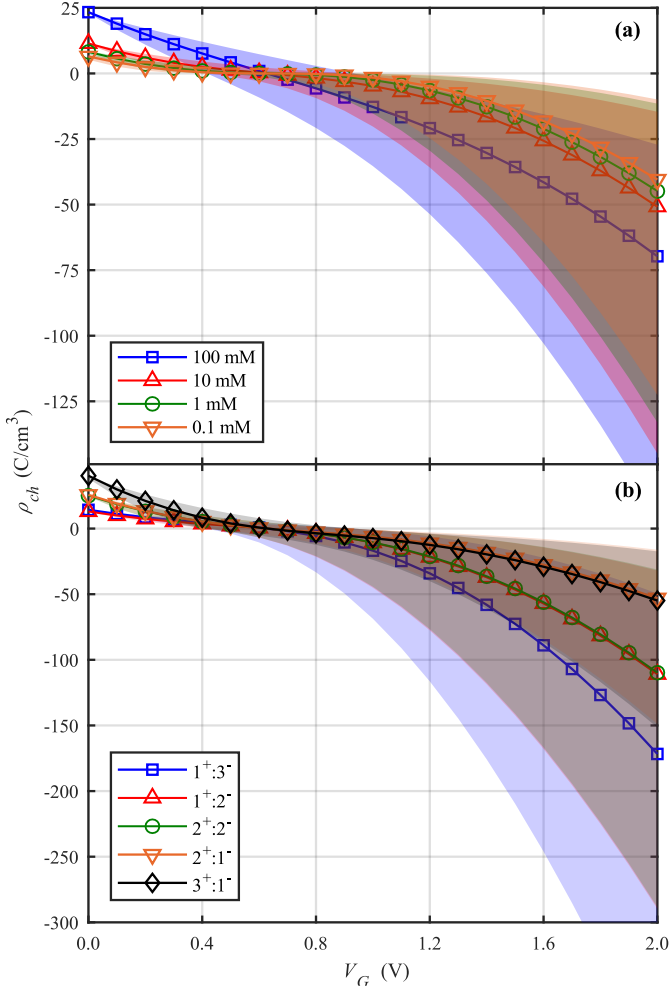

**Figure 2.** Variation of $\rho_{ch}$ versus $V_G$ for (a) a monovalent symmetrical electrolyte at different concentrations and (b) a 10 mM solution with different electrolyte symmetries. The average (symbols) indicates the value of $\rho_{ch}$ at $D_{ave}$ and the shaded area depicts values at $D_{min}$ and $D_{max}$ of the channel.

Increase in $V_G$ above 0.9 V shifts $\rho_{ch}$ at the entire channel walls toward negative values. The possibility of charge inversion obtained here is called the ambipolar effect. For the devices with planar gates (using metallic strips deposited on top of the channel), the experimental results of Karnik *et al.* [33] indicate ambipolar effect may not be observed for oxides with a low dielectric constant such as $SiO_2$. However, Guan *et al.* [28] pointed out the charge inversion may occur at relatively higher voltages in which the leakage current problem appears. Later, Lee *et al.* [34] showed that a surrounded gate structure facilities the ambipolar effect. Thus, the ambipolarity of a fluidic FET device strongly depends on the gate structure. Besides, the converging-diverging geometry promotes charge inversion at a relatively lower $V_G$ as compared to those reported previously [28] and consequently avoids problems arising from leakage current. However, this geometry does not allow generating a uniform negative charge. Thus, the difference of $\rho_{ch}$ at $D_{min}$ and $D_{max}$ (indicated by the width of the shaded area) increases at very high $V_G$. Employing higher concentrations or anion with a higher valance (e.g. $1^+:3^-$), a negative $\rho_{ch}$ is achieved at lower $V_G$ and increases more rapidly with an increase in $V_G$. It is also worth noting that at zero $V_G$ the cation valance is dominating for the obtained $\sigma_0$ e.g. $1^+:1^-$, $1^+:2^-$ and $1^+:3^-$ electrolytes present the same $\sigma_0$. While at high $V_G$, the anion valance becomes dominating e.g. $1^+:2^-$ and $2^+:2^-$ have the same $\rho_{ch}$ at $V_G > 0.7$ V. This can be explained by the fact that $\sigma_0$ is achieved in the response of hydroxyl groups at channel walls those are mostly negatively charged. Thus, the valance and concentration of cations determine $\rho_{ch}$ in fluid next to channel walls. On the other hand, a positive $V_G$ enables depletions of cation and attraction of anions next to channel walls. As a result, the valance and concentration of anions become dominating in the obtained $\rho_{ch}$ in fluid next to channel walls.

*3.2. Zeta potential*
**Figure 3** shows the variation of the $\zeta$ potential with $V_G$ at different electrolyte concentrations and symmetries. The average (indicted by symbols), minimum and maximum (shaded areas) are $\zeta$ potential values at $D_{ave}$, $D_{min}$ and $D_{max}$, respectively. In all cases, applying a $V_G > 0$ shifts $\zeta$ potential towards positive values. It is also worth noting that $\zeta$ potential at $D_{max}$ is always highest when compared to the rest of the channel. Almost all data sets depict a non-linear behavior of $\zeta$ potential with applied $V_G$ [35]. We have also developed a simple analytical model presented in the appendix which can predict both linear and non-linear behavior depending on the capacitance of the diffusion layer. Thus, the linear behavior of 100 mM concentration arises from the fact that the diffuse layer remains unchanged with applying $V_G$ according to equation (A9). A similar linear to non-linear transition due to dilution of the electrolyte has been demonstrated earlier using density functional theory and Monte Carlo simulations [36]. However, in the latter study, the variation of $\zeta$ potential was considered as a function of surface charge density while here we change surface charge through applying $V_G$. Using analytical methods and including the Stern layer, a linear to non-linear trend is observed with increasing pH in a nanoslit with and without gate [37-39].

At zero $V_G$, the variation of both electrolyte concentration and symmetry enables modulation of $\zeta$ potential. However, positive $\zeta$ potential is not achieved by increasing the monovalent electrolyte concentration (cf. figure 3a). While a high valance cation ($3^+:1^-$) gives a positive $\zeta$ potential as can be seen in figure 3b. Nevertheless, neither concentration nor symmetry is as efficient as applying $V_G$ for variation of $\zeta$ potential. It is also worth noting that dilution of electrolyte makes it more sensitive to the variation of $V_G$ and *vice versa*. This can be explained by the increased Debye length at lower concentrations which allows $V_G$ to manipulate ion distribution further away from the channel walls. For different electrolyte symmetries, the variation of $\zeta$ potential with the $V_G$ is the most efficient for $1^+:1^-$ and $2^+:1^-$, less for $1^+:2^-$, $3^+:1^-$ and least for $1^+:3^-$ and $2^+:2^-$ electrolyte. Thus, in general, a higher valance anion reduces the efficiency of $\zeta$ potential modulation by $V_G$. This can be partially explained by the fact that a $1^+:3^-$ electrolyte has three times more cation concentration than a monovalent counterpart and higher concentration shows a smaller variation of $\zeta$ potential with $V_G$ as can be seen in figure 3a.

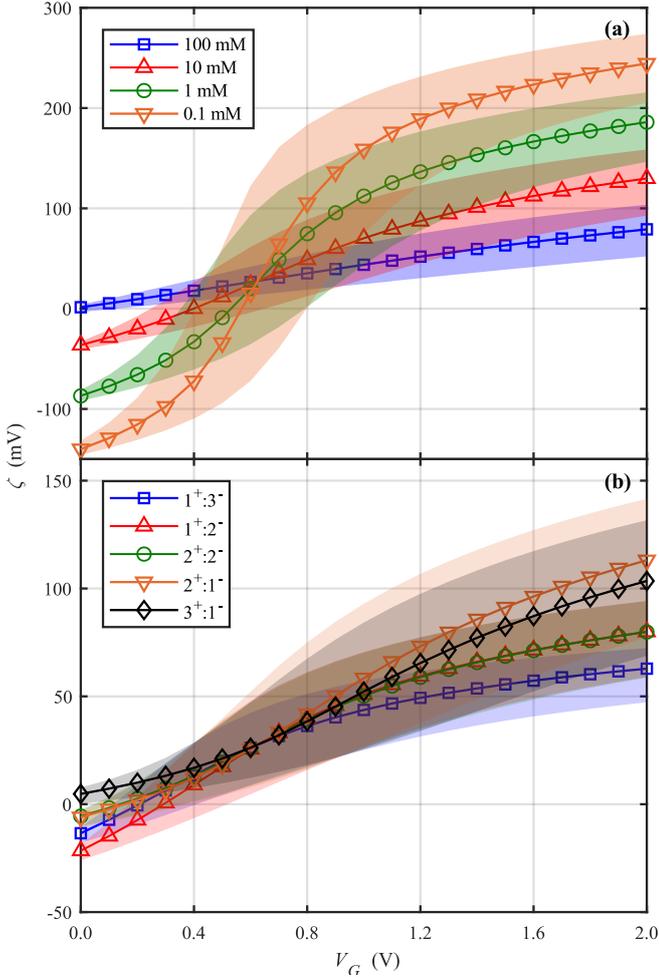

**Figure 3.** Variation of $\zeta$ potential with $V_G$ for (a) a monovalent symmetrical electrolyte at different concentrations and (b) a 10 mM solution with different electrolyte symmetries. The average (symbol), maximum and minimum (shaded area) are denoting values at $D_{max}$, $D_{ave}$, and $D_{min}$ in the channel.

According to the Helmholtz-Smoluchowski equation ($\overline{u} = -\varepsilon E \overline{\zeta}/\eta$), $\overline{\zeta}$ is the average $\zeta$ potential in the entire channel which determines the net flow of the channel [4-6, 40]. For the 100 mM monovalent electrolyte, the $\overline{\zeta}$ is nearly zero at $V_G = 0$ meaning that the diffused layer thickness and consequently electroosmotic flow is negligible. For each lower concentration, however, $\overline{\zeta} = 0$ is achieved at a specific $V_G$ which means there is a circulation of fluid inside the channel although the net flow is zero [27]. Since $\overline{\zeta} = 0$ was achieved via applying $V_G$, one can overcome the problem of low flow rates by employing a pulsed $V_G$.

Furthermore, the difference in the maximum and minimum $\zeta$ potential ($\Delta\zeta$) is shown in **Figure 4**. In the absence of an applied $V_G$, there is little variation in $\Delta\zeta$ when concentration is varied (cf. figure 4a). This results from the channel curvature, along with the longitudinal field, that generates a nonuniform $\zeta$ potential. Thus, dilution of the electrolyte increases the molar conductivity of the fluid and consequently increases the field gradient at $D_{min}$ which is detected as a slight increase in $\Delta\zeta$.

The dashed line in figure 4a indicates the maximum $\Delta\zeta$ ($\Delta\zeta_{max}$) which depends on the concentration. Thus, a diluted electrolyte is preferred if a higher $\Delta\zeta_{max}$ is required. Besides, $\Delta\zeta_{max}$ is achieved with a lower $V_G$ for a diluted electrolyte. Thus, in the devices with leakage current issue, utilizing a diluted electrolyte leads to smaller leakage current. For the various electrolyte symmetries, the $\Delta\zeta$ is constant at $V_G = 0$. The $\Delta\zeta_{max}$ appears to be more dependent on the anion valance than the cation valance. As the valance of anion decreases, the $\Delta\zeta_{max}$ increases significantly e.g. $3^+:1^-$ and $2^+:1^-$ have the highest while $1^+:3^-$ has the lowest $\Delta\zeta_{max}$. Also, a lower anion valance requires increased $V_G$ to achieve $\Delta\zeta_{max}$. However, there is still a minor dependency on the valance of cation i.e. a higher valance cation increases anion concentration which lowers $\Delta\zeta_{max}$ and shifts it to a higher $V_G$.

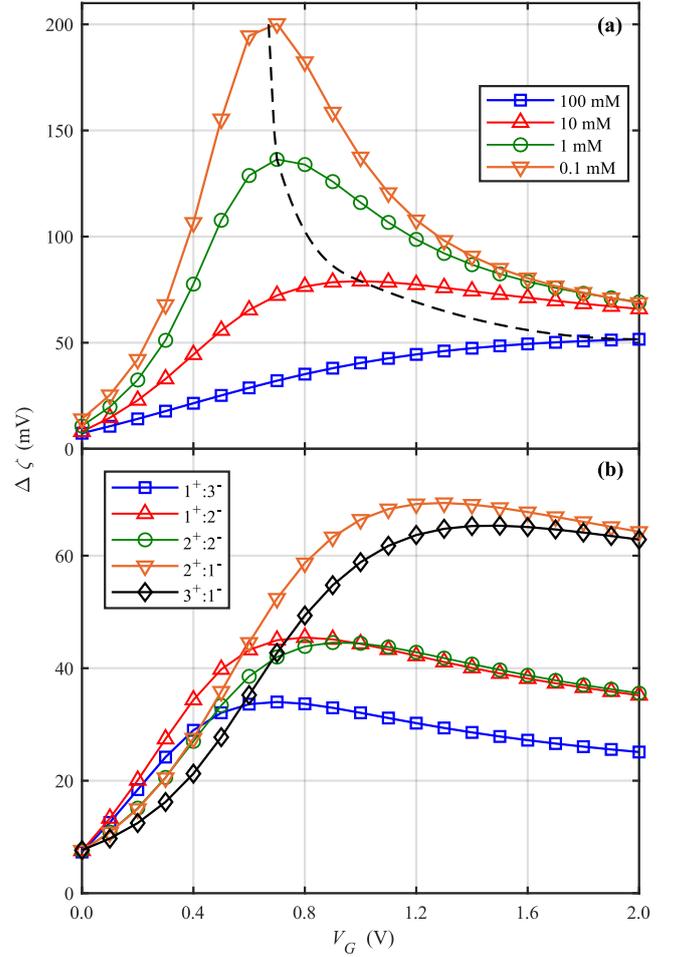

**Figure 4.** Variation of $\Delta\zeta$ with applied $V_G$ for (a) a monovalent symmetrical electrolyte at different concentrations and (b) a 10 mM solution with different electrolyte symmetries.

*3.3. Velocity field*
**Figure 5** shows the axial velocity, $U_l$, profile at $D_{max}$ for a range of $V_G$ (0-2 V). It can be seen that $U_l$ depends on $V_G$ as well as the electrolyte's concentration and symmetry. However, it appears that $V_G$ is more effective for example in reversing the flow direction at $D_{max}$ while the change in electrolyte concentration and symmetry does not lead to a reversal of flow direction. Another interesting feature in the figure is the reversal of the flow direction next to the wall as opposed to compared to channel center ($r = 0$) at 0.8 V which is indeed an indicator of flow circulation.

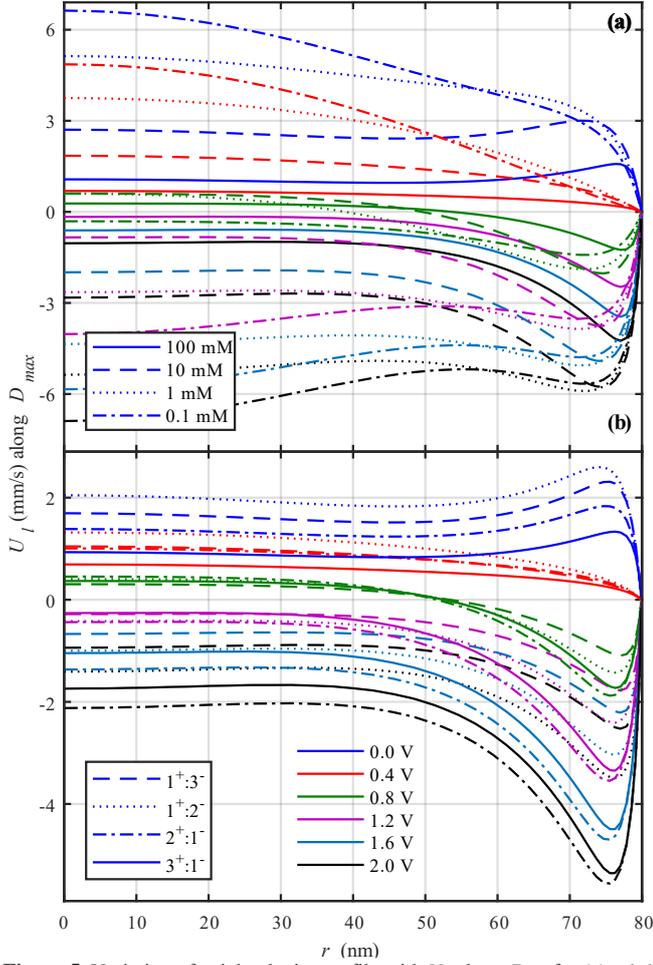

**Figure 5.** Variation of axial velocity profile with $V_G$ along $D_{max}$ for (a) a 1-1 symmetrical electrolyte at different concentrations and (b) a 0.01 M solution with different nonsymmetrical electrolytes.

To provide more details on the flow circulation process, we plotted electroosmotic velocity in the side view cross-section of the channel in **Figure 6** at $V_G$ of 0.8 V. As required by continuity condition, at $V_G = 0$ the velocity is maximum in the converging sections (cf. S1-5 in supplementary material). Thus, a moving particle experiences acceleration and deceleration while passing through the channel. For $V_G = 0.8$ V the velocity is maximum next to the $D_{max}$ channel walls. The minimum flow rate is obtained at 100 mM. There is also a vortex below $D_{max}$ that can be used for particle trapping as illustrated in our previous work [27]. As the concentration decreases, a net flow is developed to the left which also shifts the vortex towards centerline ($r = 0$), and eventually, for 0.1 mM, there is no possibility of circulation and trapping. Variation of the velocity field is negligible for the different electrolyte symmetries. It is worth noting that while it is possible to generate vortex with any electrolyte condition using proper $V_G$ but the available window for each electrolyte and symmetry may differ (cf. S1-5 in supplementary material).

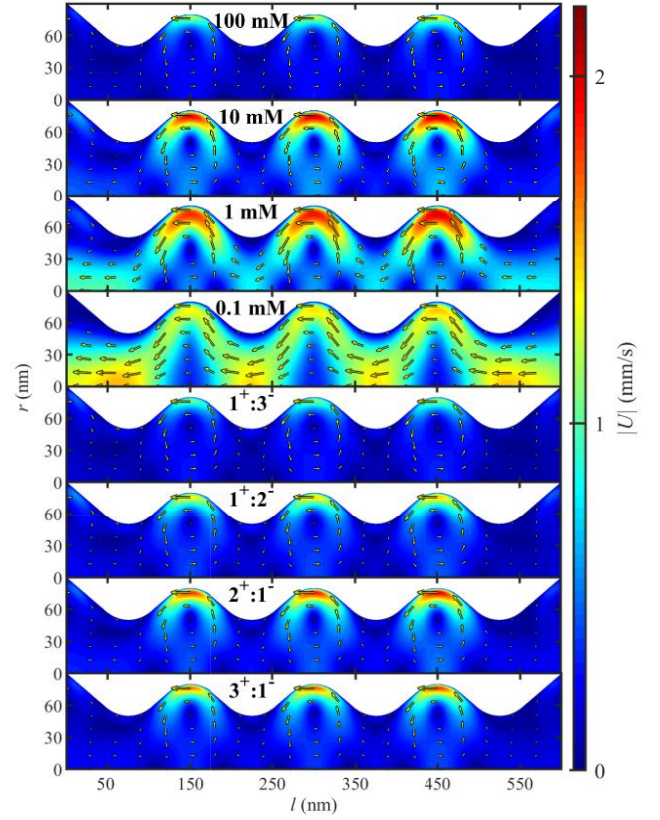

**Figure 6.** Variation of the velocity field at $V_G = 0.8$ V for 1:1 electrolyte at different concentrations and a 10 mM solution with different electrolyte symmetries. The arrows and color are indicating the flow direction and magnitude of velocity, respectively.

## 4. Conclusion

In conclusion, a converging-diverging geometry is feasible by taking advantage of the sequential nature of DRIE which remains intact after oxidation. The field-effect control was achieved by applying 0-2 V to gate electrode, i.e. unoxidized Si that surrounded the outer surface of $SiO_2$ channel walls, which impose radial electric field and effectively modulate the surface charge and $\zeta$ potential at the $SiO_2$/fluid interface. The results show applying $V_G > 0$ generates heterogeneous surface charge in diverging and converging sections and even enables the ambipolar effect. The charge inversion was achieved at much lower $V_G$ compared to previously reported devices and thus eliminating the leakage current problem. The $\zeta$ potential as a function of $V_G$ shows a linear to non-linear transition by dilution of electrolyte in agreement with density functional theory and Monte Carlo simulations. The difference in $\zeta$ potential across the channel ($\Delta\zeta$) considerably increases by dilution of the electrolyte. Monovalent symmetric electrolyte results in higher $\Delta\zeta$ compared to asymmetric ones and anion valance was found to have a major effect compared to cations valance. The result shows that $\Delta\zeta$ presents a maximum with the $V_G$ which has not been reported so far. The proper $V_G$ for $\Delta\zeta_{max}$ decreases with both dilution of electrolyte and higher anion valance which is important to overcome the leakage current problem in fluidic FETs. While the velocity field can be altered by changing electrolyte concentration and symmetry, applying $V_G$ was found to be a more efficient way than electrolyte modifications. For instance, generating a circulation inside the channel is only achieved by proper $V_G$ which on its own can be exploited for efficient mixing. This is very promising since most mixer designs introduced to date needed complicated geometries, at least two adjacent gate electrodes. The present design alleviates the need for multi-electrodes and requires a simple fabrication technique.

## 5. Acknowledgment

We thank H. van Wolferen and J. Bomer respectively from Nanolab and BIOS in Mesa+ Institute for Nanotechnology, who shared their expertise in nanofabrication. This work was partially supported by the Iranian Ministry of Science, Research and Technology travel grant, and the Iranian Nanotechnology Initiative Council.

**Declarations of interest:** none

## Appendix

Assuming that the EDL and gate oxide are parallel planes the Gauss' law gives:

$$\varepsilon_f E_r^f - \varepsilon_{ox} E_r^{ox} = \frac{\sigma}{\varepsilon_0} \qquad (5)$$

here, $ox$ and $f$ superscripts denote radial electric field, $E_r$, in the gate oxide and fluid respectively. $\varepsilon_0$ is the vacuum dielectric constant, $\varepsilon_{ox}$ and $\varepsilon_f$ are relative permittivities of the gate oxide and fluid, respectively. $\sigma$ is the surface charge density in the fluid adjacent to the oxide.

Assuming that the potential in the bulk is zero, the equation (5) can be rewritten as,

$$C_d \phi - C_{ox}(V_G - \phi) = \sigma \qquad (6)$$

where $C_{ox} = \varepsilon_0 \varepsilon_{ox}/t_{ox}$ is the capacitance per unit area of the oxide layer with thickness $t_{ox}$, and $C_d = d\sigma/d\phi$ is the differential capacitance. As illustrated by Grahame [41], the Gouy-Chapman-Stern model divides $C_d$ to the capacitance of the charges held at the outer Helmholtz layer ($C_H$) and capacitance of the truly diffused charge ($C_{diff}$). The former is independent of $\phi$, but the latter varies in a v-shaped fashion with the $\phi$. The composite capacitance shows a complex behavior and is governed by the smaller of the two components. At larger electrolyte concentrations, or even at large polarizations in dilute media, $C_{diff}$ becomes so large that it no longer contributes to $C_d$ and one sees only the constant capacitance of $C_H$. However, in general, $C_{diff}$ is smaller and shows a major contribution to the total capacitance in series. Here, the EDL capacitance is assumed to be equal to that of the diffused layer.

$$\zeta = \frac{\sigma + C_{ox}V_G}{(C_{ox} + C_{diff})} \qquad (7)$$

When there is no voltage on the gate:

$$\zeta_0 = \frac{\sigma}{(C_{ox} + C_{diff})} \qquad (8)$$

where $\zeta_0$ is the intrinsic $\zeta$ potential of the channel wall. Now rewriting equation (7) gives:

$$\zeta = \frac{C_{ox}}{(C_{ox} + C_{diff})} V_G + \zeta_0 \qquad (9)$$

Since $C_{diff}$ is a function of $V_G$, the equation (9) does not predict a linear variation of $\zeta$ potential with $V_G$. However, at very high electrolyte concentration $C_{diff}$ shows very limited change due to applied $V_G$ and thus $\zeta$ potential increase linearly with $V_G$. Thus this simplified model explains the variation of $\zeta$ potential with the different electrolyte concentrations and symmetries in figure 3.

## References

1. Li, D., *Electrokinetics in microfluidics*. Vol. 2. 2004: Academic Press.
2. Kang, Y. and D. Li, *Electrokinetic motion of particles and cells in microchannels.* Microfluidics and nanofluidics, 2009. **6**(4): p. 431-460.
3. Herr, A.E., et al., *Electroosmotic Capillary Flow with Nonuniform Zeta Potential.* Analytical Chemistry, 2000. **72**(5): p. 1053-1057.
4. Anderson, J.L. and W. Keith Idol, *Electroosmosis through pores with nonuniformly charged walls.* Chemical Engineering Communications, 1985. **38**(3-6): p. 93-106.
5. Ajdari, A., *Electro-osmosis on inhomogeneously charged surfaces.* Physical Review Letters, 1995. **75**(4): p. 755.
6. Ajdari, A., *Generation of transverse fluid currents and forces by an electric field: electro-osmosis on charge-modulated and undulated surfaces.* Physical Review E, 1996. **53**(5): p. 4996.
7. Erickson, D. and D. Li, *Influence of Surface Heterogeneity on Electrokinetically Driven Microfluidic Mixing.* Langmuir, 2002. **18**(5): p. 1883-1892.
8. Tian, F., B. Li, and D.Y. Kwok, *Tradeoff between Mixing and Transport for Electroosmotic Flow in Heterogeneous Microchannels with Nonuniform Surface Potentials.* Langmuir, 2005. **21**(3): p. 1126-1131.
9. Fu, L.-M., J.-Y. Lin, and R.-J. Yang, *Analysis of electroosmotic flow with step change in zeta potential.* Journal of Colloid and Interface Science, 2003. **258**(2): p. 266-275.
10. Qian, S. and H.H. Bau, *A Chaotic Electroosmotic Stirrer.* Analytical Chemistry, 2002. **74**(15): p. 3616-3625.
11. Ghowsi, K. and R.J. Gale, *Field-effect electroosmosis.* Journal of Chromatography A, 1991. **559**(1-2): p. 95-101.
12. Hayes, M.A. and A.G. Ewing, *Electroosmotic flow control and monitoring with an applied radial voltage for capillary zone electrophoresis.* Analytical Chemistry, 1992. **64**(5): p. 512-516.
13. Jiang, Z. and D. Stein, *Charge regulation in nanopore ionic field-effect transistors.* Physical Review E, 2011. **83**(3): p. 031203.
14. Daiguji, H., P. Yang, and A. Majumdar, *Ion transport in nanofluidic channels.* Nano Letters, 2004. **4**(1): p. 137-142.
15. Ai, Y., et al., *Field-effect regulation of DNA translocation through a nanopore.* Analytical chemistry, 2010. **82**(19): p. 8217-8225.
16. Ai, Y., et al., *Ionic current rectification in a conical nanofluidic field-effect transistor.* Sensors and Actuators B: Chemical, 2011. **157**(2): p. 742-751.
17. He, Y., et al., *Gate manipulation of DNA capture into nanopores.* Acs Nano, 2011. **5**(10): p. 8391-8397.

18. He, Y., et al., *Controlling DNA translocation through gate modulation of nanopore wall surface charges.* ACS nano, 2011. **5**(7): p. 5509-5518.
19. Jin, X. and N. Aluru, *Gated transport in nanofluidic devices.* Microfluidics and nanofluidics, 2011. **11**(3): p. 297-306.
20. Singh, K.P. and M. Kumar, *Effect of gate length and dielectric thickness on ion and fluid transport in a fluidic nanochannel.* Lab on a Chip, 2012. **12**(7): p. 1332-1339.
21. Matovic, J., et al., *Field-effect transistor based on ions as charge carriers.* Sensors and Actuators B: Chemical, 2012. **170**: p. 137-142.
22. Xue, S., N. Hu, and S. Qian, *Tuning surface charge property by floating gate field-effect transistor.* Journal of colloid and interface science, 2012. **365**(1): p. 326-328.
23. Benson, L., et al., *Field-effect regulation of Donnan potential and electrokinetic flow in a functionalized soft nanochannel.* Soft Matter, 2013. **9**(41): p. 9767-9773.
24. Pardon, G. and W. van der Wijngaart, *Modeling and simulation of electrostatically gated nanochannels.* Advances in colloid and interface science, 2013. **199**: p. 78-94.
25. Yeh, L.-H., et al., *Field-effect control of surface charge property and electroosmotic flow in nanofluidics.* The Journal of Physical Chemistry C, 2012. **116**(6): p. 4209-4216.
26. Hughes, C., L.-H. Yeh, and S. Qian, *Field-effect modulation of surface charge property and electroosmotic flow in a nanochannel: Stern layer effect.* The Journal of Physical Chemistry C, 2013. **117**(18): p. 9322-9331.
27. Kateb, M., M. Kolahdouz, and M. Fathipour, *Modulation of heterogeneous surface charge and flow pattern in electrically gated converging-diverging nanochannel.* International Communications in Heat and Mass Transfer, 2018. **91**: p. 103-108.
28. Guan, W., R. Fan, and M.A. Reed, *Field-effect reconfigurable nanofluidic ionic diodes.* Nature communications, 2011. **2**: p. 506.
29. Schasfoort, R.B.M., et al., *Field-Effect Flow Control for Microfabricated Fluidic Networks.* Science, 1999. **286**(5441): p. 942.
30. Wolferen, H. and L. Abelmann, *Laser interference lithography*, in *Lithography: Principles, Processes and Materials*, T.C. Hennessy, Editor. 2011, NOVA Publishers: Hauppauge NY, USA. p. 133-148.
31. Jing, D. and B. Bhushan, *The coupling of surface charge and boundary slip at the solid–liquid interface and their combined effect on fluid drag: A review.* Journal of Colloid and Interface Science, 2015. **454**: p. 152-179.
32. Hong, S., et al. *Numerical study of mixing in microchannels with patterned zeta potential surfaces.* in *ASME 2003 International Mechanical Engineering Congress and Exposition.* 2003. American Society of Mechanical Engineers.
33. Bhattacharyya, S. and A.K. Nayak, *Electroosmotic flow in micro/nanochannels with surface potential heterogeneity: An analysis through the Nernst–Planck model with convection effect.* Colloids and Surfaces A: Physicochemical and Engineering Aspects, 2009. **339**(1–3): p. 167-177.
34. Lee, S.-H., et al., *Sub-10 nm transparent all-around-gated ambipolar ionic field-effect transistor.* Nanoscale, 2015. **7**(3): p. 936-946.
35. Hu, N., Y. Ai, and S. Qian, *Field-effect control of electrokinetic transport in micro/nanofluidics.* Sensors and Actuators B: Chemical, 2012. **161**(1): p. 1150-1167.
36. Yu, Y.-X., J. Wu, and G.-H. Gao, *Density-functional theory of spherical electric double layers and ζ potentials of colloidal particles in restricted-primitive-model electrolyte solutions.* The Journal of chemical physics, 2004. **120**(15): p. 7223-7233.
37. Yeh, L.-H., et al., *Gate manipulation of ionic conductance in a nanochannel with overlapped electric double layers.* Sensors and Actuators B: Chemical, 2015. **215**: p. 266-271.
38. Ma, Y., et al., *pH-regulated ionic conductance in a nanochannel with overlapped electric double layers.* Analytical chemistry, 2015. **87**(8): p. 4508-4514.
39. Ma, Y., et al., *Analytical model for surface-charge-governed nanochannel conductance.* Sensors and Actuators B: Chemical, 2017. **247**: p. 697-705.
40. Ren, L. and D. Li, *Electroosmotic Flow in Heterogeneous Microchannels.* Journal of Colloid and Interface Science, 2001. **243**(1): p. 255-261.
41. Grahame, D.C., *The Electrical Double Layer and the Theory of Electrocapillarity.* Chemical Reviews, 1947. **41**(3): p. 441-501.